\documentclass[12pt,onecolumn]{IEEEtran}
\usepackage{graphics, graphicx}
\usepackage{amssymb,amsmath}
\usepackage{amsmath}
\usepackage[rm]{subfigure}    
\usepackage{threeparttable}  
\usepackage[dvips]{color}
\usepackage{ams fonts}  
\usepackage{url}   
\usepackage[latin1]{inputenc} 
\usepackage{pifont}
\usepackage{cite}
\usepackage[dvips]{color}
\usepackage{lineno}
\usepackage{pstricks}
\begin{document}
\title{Power-Rate Allocation in DS/CDMA Based on Discretized Verhulst Equilibrium}
\author{\IEEEauthorblockN{Lucas Dias H. Sampaio$^\S$, Moisés F.Lima$^\S$, Mario Lemes Proença Jr.$^\S$ \& Taufik Abrão$^{\ddag,\,\S}$}\\
\vspace{4mm}
\IEEEauthorblockA{\small $^\S$Computer Department, State University of Londrina, PR, 86051-990, Brazil\\
\texttt{\small lucas.dias.sampaio@gmail.com, \,\, mflima@uel.br, \,\, proenca@uel.br, \,\, taufik@uel.br}}\\
\IEEEauthorblockA{\small $^\ddag$Dept. of Electrical Engineering; State University of Londrina, Brazil; \\ 
\texttt{\small taufik@uel.br\quad www.uel.br/pessoal/taufik}}}

\maketitle

\begin{abstract}
This paper proposes to extend the discrete Verhulst power equilibrium approach, previously suggested in \cite{Gross2010}, to the power-rate optimal allocation problem. Multirate users associated to different types of traffic are aggregated to distinct user' classes, with the assurance of minimum rate allocation per user and QoS. Herein, Verhulst power allocation algorithm was adapted to the single-input-single-output DS/CDMA jointly power-rate control problem. The analysis was carried out taking into account the convergence time, quality of solution, in terms of the normalized squared error (NSE), when compared with the analytical solution based on interference matrix inverse, and computational complexity. Numerical results demonstrate the validity of the proposed resource allocation methodology.
\end{abstract}


\section{Introduction} \label{sec:INTRO}
In the last years many efforts has been spent trying to find the best resource allocation algorithm that could be easy applied to DS/CDMA communications systems. The Foschini and Miljanic \cite{Foschini_93} studies can be considered as a foundation of many well-known distributed power control algorithms (DPCA) in scientific literature, because they try to solve an ordinary differential equation (ODE, eq. (1) in \cite{Foschini_93}), which with some minor alterations is also considered in many other subsequent studies. Therefore, a new ODE can lead to a new algorithm, more promising in several aspects, such as, convergence, proximity to the optimum value, and sensibility to estimation errors as well.

With this context in mind, the work in \cite{Gross2010} proposed and analysed a new ODE for the DPCA based on the Verhulst equation \cite{Verhulst1838}. The discrete version of the Verhulst population model is more diffused in the literature and it is called logistic map. The logistic map was studied thoroughly by R.M. May in \cite{May76}. Von Neumann and Ulam \cite{Ulam47} also studied the logistic map and they evaluated the possibility to use it as random generator number, which is gotten in certain conditions.

The Verhulst model was initially designed to describe population growth of biological species with food and physical space restriction. With the successfully mathematical model adaptation to power control in a single-rate DS/CDMA systems proposed in \cite{Gross2010}, this work suggests an expansion of the Verhulst approach to other optimization problems, such as the power-rate allocation problem, and its multi-objective versions, due to DPCA performance $\times$ complexity gain when compared to the classical algorithms such as Foschini or sigmoidal.

In this paper we have adapted the Verhulst approach to the power-rate allocation problem with multirate QoS associated to different types of traffic (based on user classes), and satisfying the minimum rate allocation per user requirement. Hence, the Verhulst power allocation algorithm of \cite{Gross2010} was adapted  to the power-multirate control problem. The analysis was carried out taking into account the convergence time, quality of solution when compared with the analytical solution based on interference matrix inverse, and computational complexity.

This paper is organized as follows: Section \ref{sec:Classical} gives an overview of the power control classical solution and how it is adapted to multirate problem. In Section \ref{sec:Verh_algo} the power-rate Verhulst algorithm is proposed. Numerical results with corresponding simulation parameters setup are treated in Section \ref{sec:NR}. Finally, the conclusions are offered in Section \ref{sec:concl}.

\section{Power and Rate Allocation Problem} \label{sec:Classical}
In a multiple access system, such as direct sequence code division multiple access (DS/CDMA), the power control problem is of great importance in order to achieve relevant system capacity and throughput. The power control problem can be solved by a vector that contain the minimum power to be assigned in the next time slot to each active user, in order to achieve the minimum quality of service (QoS) through the minimum carrier to interference ratio (CIR).

In multirate multiple access wireless communications systems the bit error rate (BER) is often used  as a QoS measure and, since the BER is directly linked to the signal to interference plus noise ratio (SNIR), we are able to use the SNIR parameter as QoS measurement. Hence, associating the SNIR to the CIR at time slot $n$ results:
\begin{equation}\label{SNR_CIR}
 \delta_i[n] = \dfrac{R_c}{R_i[n]} \times \Gamma_i[n], \qquad n = 0, 1, \ldots N
\end{equation}
where $\delta_i[n]$ is the SNIR of user $i$ at the $n$th iteration, $R_c$ is the chip rate, $R_i[n]$ is the data rate for user $i$, $\Gamma_i[n]$ is the CIR for user $i$ at iteration $n$, and $N$ is the maximal number of iterations. From (\ref{SNR_CIR}) we are able to calculate the data rate for user $i$ at iteration $n$:
\begin{equation}\label{RATE}
 R_i[n] = \dfrac{R_c}{\delta_i[n]} \times \Gamma_i[n], \qquad n = 0, 1, \ldots, N
\end{equation}

The CIR for the $i$th user can be calculated as \cite{Gross2010, Elmusrati_03}:
\begin{equation}\label{CIR}
 \Gamma_i[n] = \dfrac{P_i[n] g_{ii}[n]}{\sum\limits_{\scriptstyle j=1\hfill \atop \scriptstyle j\ne i\hfill}^{K} P_i[n]g_{ij}[n] + \sigma^2}, i = 1,\ldots,K
\end{equation}
where $P_i[n]$ is the power allocated to the $i$th user at time slot $n$ and is bounded by $[P_{\min};\,P_{\max}]$, the channel gain (including path loss, fading and shadowing effects) between user $j$ and user (or base station) $i$ is identified by $g_{ij}$, $K$ is the number of active users in the system, and $\sigma_i^2=\sigma_j^2=\sigma^2$ is the average power of the additive white Gaussian noise (AWGN) at the input of $i$th receiver, admitted identical for all users. Therefore, in DS/CDMA multirate systems the CIR relation to achieve the minimum rate can be calculated to each user class as follows \cite{Elmusrati_08}:
\begin{equation}\label{CIR_multipleClass}
 \Gamma^{\ell}_{\min} = \dfrac{R_{\min}^{\ell} \delta^*}{R_c}, \qquad \ell=1\cdots L
\end{equation}
where $\Gamma^{\ell}_{\min}$ and $R_{\min}^{\ell}$ is the minimum CIR and minimum user rate associated to the $\ell$th user class, respectively, $\delta^*$ is the minimum (or target) signal to noise ratio (SNR) to achieve minimum acceptable BER (or QoS), and $L$ is the total number of user classes in the system (voice, data, video, and so on). Besides, the power allocated to the $k$th user belonging to the $\ell$th class at $n$th iteration is:
\begin{equation}\label{POWER2}
 p_{k}^{\ell}[n], \quad k=1\cdots K_{\ell}; \quad \ell=1\cdots L,
\end{equation}
hence, the total number of active users in the system is given by $K=K_1 \cup \ldots  \cup  K_{\ell} \cup \ldots \cup K_L$. Note that indexes associated to the $K$ users are obtained by concatenation of ascending rates from different user's classes. Hence, $K_1$ identifies the lowest user's rate class, and $K_L$ the highest.

The $K\times K$ channel gain matrix, considering path loss, shadowing and fading effects, between user $j$ and user $i$ (or base station) is given by:
$$
{\bf G} =
\left[ {\begin{array}{*{20}c}
   {g_{11} } & {g_{12} } &  \cdots  & {g_{1K} }  \\
   {g_{21} } & {g_{22} } &  \cdots  & {g_{2K} }  \\
    \vdots  &  \vdots  &  \ddots  &  \vdots   \\
   {g_{K1} } & {g_{K2} } &  \cdots  & {g_{KK} }  \\
\end{array}} \right],
$$
which could be assumed static or even dynamically changing over the optimization window ($N$ time slots).

Assuming multirate user classes we are able to adapt the classical power control problem to achieve the minimum rates for each user, simply using the Shannon capacity relation between minimum CIR and minimum rate in each user class, resulting:
\begin{equation}\label{gamma_min}
\Gamma^{\ell}_{\min}=2^{R^{\ell}_{\min}}-1
\end{equation}

Now, considering a $K \times K$ interference matrix $B$
\begin{eqnarray}\label{eq:B}
{\bf B}_{ij} = &&\left\{ {\begin{array}{*{20}l}
   0, \qquad \qquad \qquad i = j; \\
   \dfrac{\Gamma_{i,\min} g_{ji}}{g_{ii}}, \qquad i \ne j;
\end{array}} \right.
\end{eqnarray}
where $\Gamma_{i,\min}$ can be obtained from (\ref{CIR_multipleClass}), taking into account each rate class requirement, and the following column vector $K \times 1$:
\begin{equation}\label{u}
{\bf u}_i = \dfrac{\Gamma_{i,\min}\sigma^2_i}{g_{ii}},
\end{equation}
we can obtain the analytical optimal power vector allocation simply by matrix inversion as:
\begin{equation}\label{classical}
 {\bf p^*} = \left({{\bf I - B}}\right)^{-1}\bf u
\end{equation}
if and only if the maximum eigenvalue of ${\bf B}$ is smaller than 1 \cite{Seneta81}; $\bf I$ is the $K \times K$ identity matrix. In this situation, the power control problem shows a feasible solution.

Herein, the classical power allocation problem is extended to incorporate multirate criterium in order to guarantee the minimum data rate per user class. Mathematically, we want to solve the following optimization problem:
\begin{eqnarray}\label{eq:Equilibrium}
\min && {\bf p}= \left[p_1^{1}\ldots p_{K_1}^{1}, \ldots, p_1^{\ell}\ldots p_{K_{\ell}}^{\ell},\ldots, p_1^{L}\ldots p_{K_{L}}^{L}\right]  \nonumber\\
\text{s.t.} &&  P_{\min}^{\ell} \leq p_{k}^{\ell} \leq P_{\max}^{\ell}   \\
&&  R^{\ell} = R^{\ell}_{\min}, \qquad \forall k\in K_{\ell}, \text{ and } \forall  \, \ell = 1, 2,\cdots L \nonumber
\end{eqnarray}

\section{Verhulst Power-Rate Optimization Approach} \label{sec:Verh_algo}
The Verhulst mathematical model was first idealized to describe population dynamics based on food and space limitation. In \cite{Gross2010} that model was adapted to single-rate DS/CDMA distributed power control using a discrete iterative convergent equation as follows:
\begin{equation}\label{eq:Verhulst}
p_i[n + 1] = \left( {1 + \alpha } \right)p_i [n] - \alpha {\rm{ }}\left[ {\frac{{\delta_i [n]}}{{\delta_i^* }}} \right]p_i[n], \,\,\, i = 1,\cdots,K
\end{equation}
where $p_i[n + 1]$ is the user $i$ power at the $n+1$ iteration, $\alpha\in(0;1]$ is the Verhulst convergence factor, $\delta_i[n]$ is the $i$th user' SNIR at iteration $n$, $\delta_i^*$ is the minimum SNR for the $i$th user that guarantee a minimum QoS in terms of performance (BER).

The recursion (\ref{eq:Verhulst}) can be effectively implemented in the $i$th mobile unit since all necessary parameters $\alpha$, the QoS level given by  $delta_i^*$, the transmitted power $p_i[n]$, except  $\delta_i[n]$, can be considered known in the mobile unit $i$. The SINR $\delta_i[n]$ can be obtained only at the correspondent base station that demodulates the signal from user $i$. In this way, the BS estimates  $\delta_i[n]$, quantizes it in a convenient number of bits, and transmits this information to the $i$th user through the direct channel. Thus, (\ref{eq:Verhulst}) depends on local parameters just allowing that the power control works in a distributed manner, i.e., each one of the $K$ links (mobile terminals to base station) carries out separately the respective power control mechanism, justifying the name distributed power control algorithm (DPCA).

Equation (\ref{eq:Verhulst}) gives a recursive power update, close to the optimal power solution after $N$ iteration. However, originally it does not consider the rate requirements in a multirate environment. In order to achieve the QoS to each user class, (\ref{eq:Verhulst}) must be adapted to reach the equilibrium $\mathop {\lim }\limits_{n \to \infty } p_i[n] = p_i^*$ when the power allocated to each user satisfies the minimum rate constraint given in (\ref{eq:Equilibrium}). Hence, the recursive equation must be rewritten considering SNIR per user class, via (\ref{SNR_CIR}), in order to incorporate multirate scenario. The minimum CIR per user class is obtained directly by (\ref{CIR_multipleClass}). In this way, considering the relation between CIR and SNIR in a multirate DS/CDMA context, we propose the equation below in order to iteratively solve optimization problem in (\ref{eq:Equilibrium}):
\begin{eqnarray}\label{newCIR}
\delta_i[n] &=& F \times \Gamma_i[n]\\
   &=& F \times \dfrac{P_i[n] g_{ii}[n]}{\sum\limits_{\scriptstyle j=1\hfill \atop \scriptstyle j\ne i\hfill}^{K} P_i[n]g_{ij}[n] + \sigma^2}, i = 1,\ldots,K \nonumber
\end{eqnarray}
where $F$ is the spreading factor per user class, given by:
\begin{equation}
 F = \dfrac{R_c}{R^{\ell}_{\min}}
\end{equation}
Note that the CIR of $i$th user at the $n$th iteration is weighted by spreading factor; so the corresponding SNIR is inversely proportional to the actual (minimum) rate of the $i$th user of the $\ell$th class.

\subsection{Quality of Solution $\times$ Convergence Speed}\label{sec:conv_speedup}
The quality of solution achieved by iterative Verhulst equation (\ref{eq:Verhulst}) is measured by how close to the optimum solution is $p[n]$, and can be quantified by means of the normalized squared error (NSE) when equilibrium is reached. The NSE definition is given by:
$$
NSE[n] = \mathbb{E}\left[ {\frac{{\left\| {{\bf{p}}[n] - {\bf{p}}^ *  } \right\|^2 }}{{\left\| {{\bf{p}}^ *  } \right\|^2 }}} \right],
$$
where $\|\cdot\|^2$ denotes the squared Euclidean distance to the origin, and $\mathbb{E}[\cdot]$ the expectation operator.

On the other hand, the convergence speed in Verhulst equation is dictated by the parameter $\alpha$. Hence, for small values of convergence factor, i.e., $\alpha\rightarrow0$, the convergence is slow, but the NSE is very small after $N$ iteration, when compared with the opposite configuration: the convergence is fast when $\alpha\rightarrow1$, but the NSE is a concern. So, in order to accelerate convergence, we propose two adaptive criteria for convergence factor $\alpha$ when iterations evolve, based on a) SNIR to target SNR difference, and b) tanh mapping for this difference, as following:
\begin{equation}\label{eq:adpt_method_a}
\hspace{-.1mm} \text{a)} \quad   \alpha_i[n] = \min\left\{\alpha_{\max};\, \frac{|\delta_i[n-1]-\delta_i^*|}{\delta_i^*} + \alpha_{\min} \right\},
\end{equation}
\begin{equation}\label{eq:adpt_method_b}
\hspace{-.1mm}  \text{b)} \quad  \alpha_i[n] = \max\left\{\alpha_{\min};\, \tanh(|\delta_i[n-1]-\delta_i^*|)\right\},
\end{equation}with $\alpha_{\min}=0.1$ and $\alpha_{\max}=0.95$.

\section{Numerical Results} \label{sec:NR}
\begin{figure}[htbp]
\centering
\includegraphics[width=.48\textwidth]{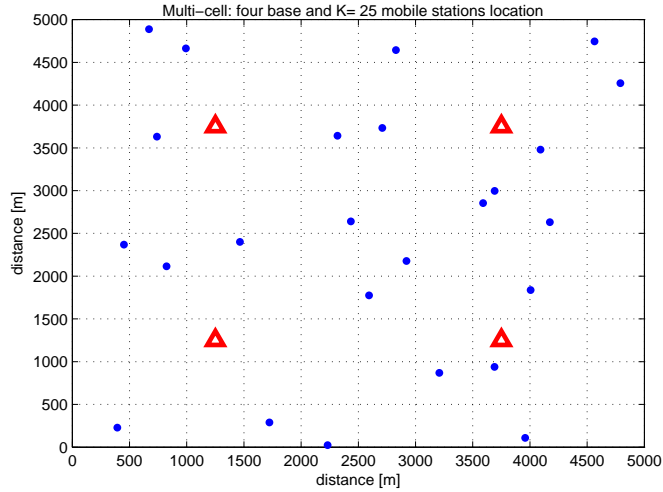}
\vspace{-4mm}
\caption{$K=25$ Mobile terminals and $4$ base stations location over 25 Km$^2$ multicell rectangular geometry.}
\label{fig:mt_bs_loc}
\end{figure}

Simulations were carried out through the MatLab ver.7.3 platform, with system parameters indicated in Table \ref{tab:param}. For all simulation results discussed in this section, it was assumed a retangular multicell geometry with a number of base station equal to $4$ and mobile terminals uniformly distributed. A typical placement for mobile terminals (mt) and base stations(BS) is provided in Fig. \ref{fig:mt_bs_loc}. Besides, the rate assignment for all multirate users was considered uniformly distributed as three submultiple rates of chip rate, $R_{\min} =[\frac{1}{128};\, \frac{1}{32};\, \frac{1}{16}]  R_c$ [bps].

A number of mobile terminals ranging from $K=5$ to $30$ was considered, which experiment slow fading channels, i.e., the following relation is always satisfied:
\begin{equation}\label{eq:Tslot_coher}
    T_{\rm slot}  <  \left(\Delta t\right)_c
\end{equation}
where $T_{\rm slot}$ is the time slot duration, and $\left(\Delta t\right)_c$  is the coherence time  of the channel\footnote{Corresponds to the time interval in which the channel characteristics do not suffer expressive variations.}. This condition is part of the SINR estimation process, and it implies that each power updating accomplished by the DPCA happens with rate of $T_{\rm slot}^{-1}$, assumed here equal to $1500$ updated per second. The recursion in (\ref{eq:Verhulst}) should converge to the optimum point before each channel gain $g_{ij}$ experiments significant changing. Note that satisfying (\ref{eq:Tslot_coher}) the gain matrices remain static during one convergence process interval.


In all simulations the entries values for the QoS targets were fixed in $\delta^*=4$ dB, the adopted receiver noise power for all users was $P_n=-63$ dBm, and the gain matrix $G$ entries had intermediate values between those used in \cite{Uykan04} and \cite{Elmusrati_08}. Furthermore, in order to evaluate and highlight the different aspects and features of the proposed resource allocation methodology, the simulation results discussed hereafter were obtained under static channels condition, situation where the channel gain' coefficients hold constant during all convergence period, i.e., for $N$ iterations executed on each $ T_{\rm slot}$ seconds. It is straightforward to show that those analyses and results can be applied considering the dynamic channels condition, where the channel coefficients changing following the coherence time of the channel, been observed the bound: $T_{\rm slot} \gg (\Delta t)_c \approx \frac{1}{f_{D_{\max}}}$, for all adopted mobilities, bounded by the maximal Doppler frequency $f_{D_{\max}}$.

\begin{table}[!htbp]
\centering
\caption{Multirate DS/CDMA system parameters} \label{tab:param}
\vspace{-.3cm}
\small
\begin{tabular}{ll}
\hline
\textbf{Parameters} & \textbf{Adopted Values}\\
\hline
\hline
\multicolumn{2}{c}{\textit{DS/CDMA Power-Rate Allocation System}} \\
\hline
Noise Power   & $P_n=-63$ [dBm]\\
Chip rate  & $R_c = 3.84\times 10^{6}$\\
Min. Signal-noise ratio& $SNR_{\min}= 4$ dB\\
Max. power per user & $P_{\max}=20$ [dBm]\\
Min. Power per user & $P_{\min}= SNR_{\min} + P_n$  [dBm]\\
Time slot duration & $T_{\rm slot} = 666.7\mu$s\\
\# mobile terminals   & $K \in \{5; \, 30\}$\\
\# base station      & BS $=4$\\
cell geometry   & rectangular, with $x_{\text{cell}}= y_{\text{cell}}= 5$ Km\\
mobile term. distrib.  & $\sim \mathcal{U}[x_{\text{cell}},\, y_{\text{cell}}]$ \\
\hline
\multicolumn{2}{c}{\textit{Channel Gain}} \\
\hline
path loss  & $\varpropto d^{-2}$ \\
shadowing & uncorrelated log-normal, $\sigma^2=6$ dB\\
fading  & Rice: $[0.6;\, 0.4]$\\
Max. Doppler freq. & $f_{D\max} = 11.1 $ Hz\\
Error estimates & $\widehat{G}=(1+\varepsilon)G,$ where $\varepsilon \sim \mathcal{U}[\pm \delta]$\\
                 & $\delta = 0: 0.02 : 0.2$\\
\hline
\multicolumn{2}{c}{\textit{User Types}} \\
\hline
\#  user classes & $L=3$ (voice, data, video)\\
User classes Rates & $R_{\min} =[\frac{1}{128};\, \frac{1}{32};\, \frac{1}{16}]  R_c$ [bps]\\
\hline
\multicolumn{2}{c}{\textit{Verhulst Power-Rate algorithm}} \\
\hline
Type & partially distributed\\
$\alpha$& range $[0.10;\, 0.95]$\\
Optimization window & $N\in [100;\,\, 1000]$ iterations\\
\hline
\multicolumn{2}{c}{\textit{Performance parameters}} \\
\hline
Trials number, TR &  $100$ samples\\
\hline
\end{tabular}
\end{table}

\subsection{Typical Convergence Performance}
\begin{figure}[htbp]
\centering
\includegraphics[width=.48\textwidth]{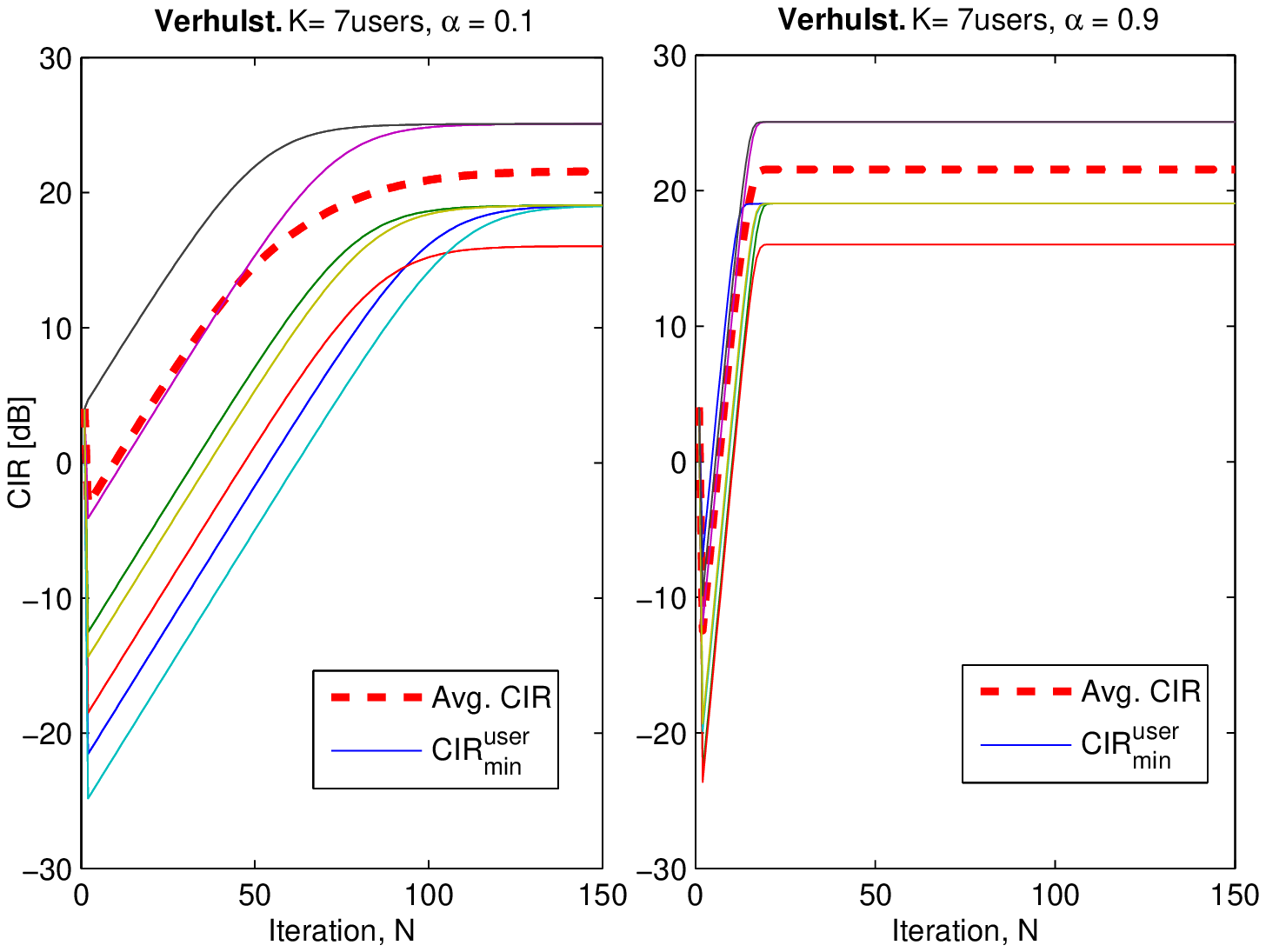}
\includegraphics[width=.48\textwidth]{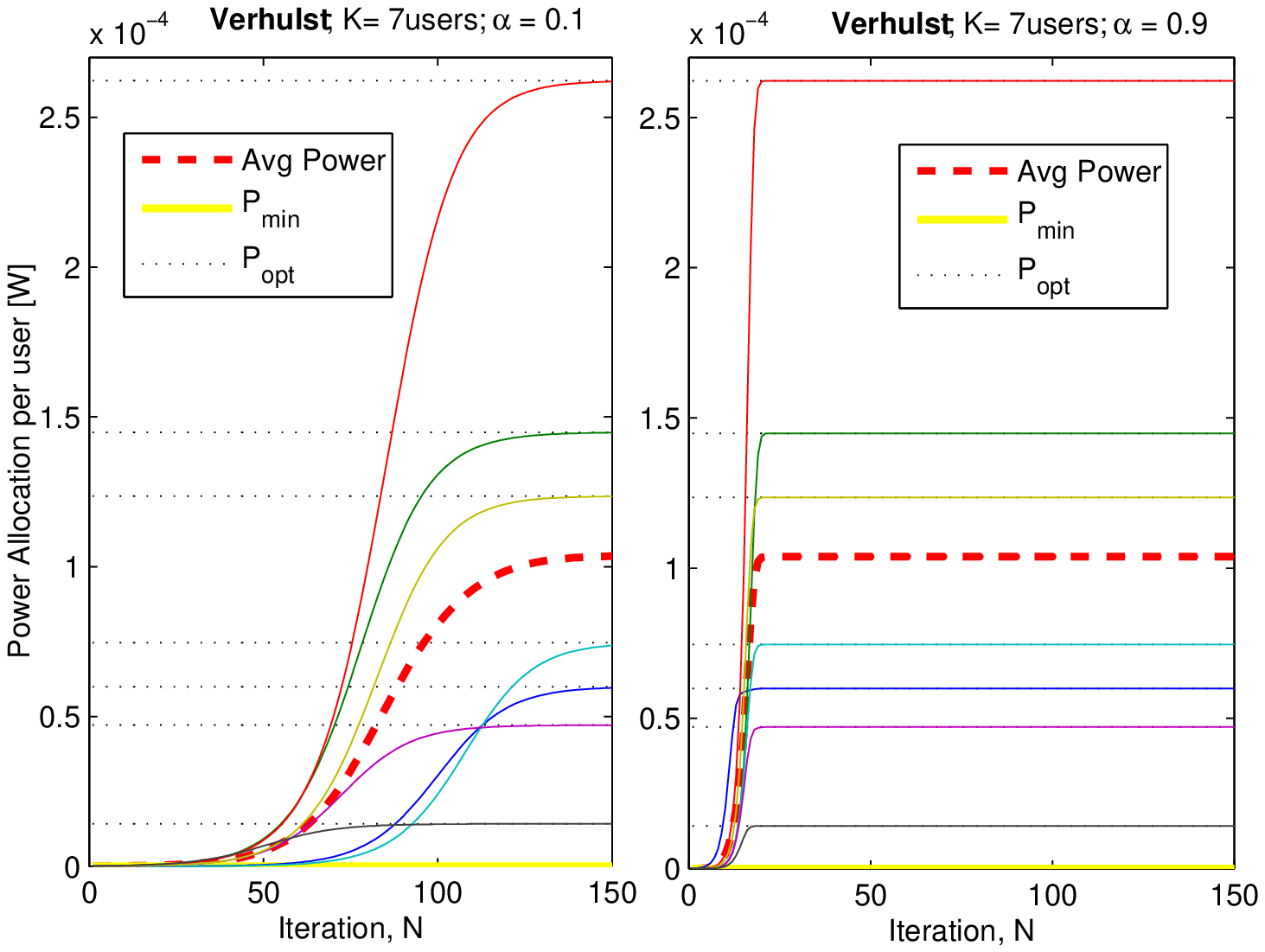}
\vspace{-4mm}
\caption{CIR and power convergence for $K=7; R_{k,min}= [120,\,  120, \, 240, \, 120, \,  30, \, 120, \,  30] [Kbps]$. $\alpha=0.1$ and $0.9$ }
\label{fig:converg_alpha0.1_0.9}
\end{figure}

Typical convergence behavior for two fixed $\alpha$ (slow and fast convergence scenarios), $K=7$ multirate users, with rate assignment uniformly distributed over the three rates, $[30;\, 120;\, 240]$ [Kbps], with $R_c = 3.84\times 10^{6}$ chips per sec, is shown in Fig.  \ref{fig:converg_alpha0.1_0.9}, for CIR (superior plots) and power allocation solution (inferior plots). Plateaux indicate convergence to the optimum power vector, $\bf p^*$; hence dot lines ($P_{\rm opt}$ in legend) indicates analytical solution given by (\ref{classical}). Note the fast convergence for all users when $\alpha=0.9$, i.e., $N\approx 25$ iterations, against $\approx 150$ for $\alpha=0.1$. Evidently, the quality of solution in this two situation is distinct, as discussed in the next subsection.

\subsection{Performance under Channel Gain Error Estimates}
In a real scenario, the SINR estimations at BS are not perfect, in the sense that the obtained values by estimation possess a random error characteristic. In order to incorporate this characteristic, a random error is added in each element of channel gain matrix, in each iteration basis. The ratio of the estimated and real channel gain values is given by $\widehat{g_{ij}}=(1+\varepsilon) g_{ij}$, where $\varepsilon$  will be considered as a random variable with uniform distribution in the range $[-\delta; \delta]$. In the subsequent simulations the adopted range values for $\delta$ were  $0$ to $0.2$, in steps of $0.02$.

\subsubsection{Dependence of Solution Quality in terms of $\alpha$}
\begin{figure}[!htbp]
\centering
\includegraphics[width=.47\textwidth]{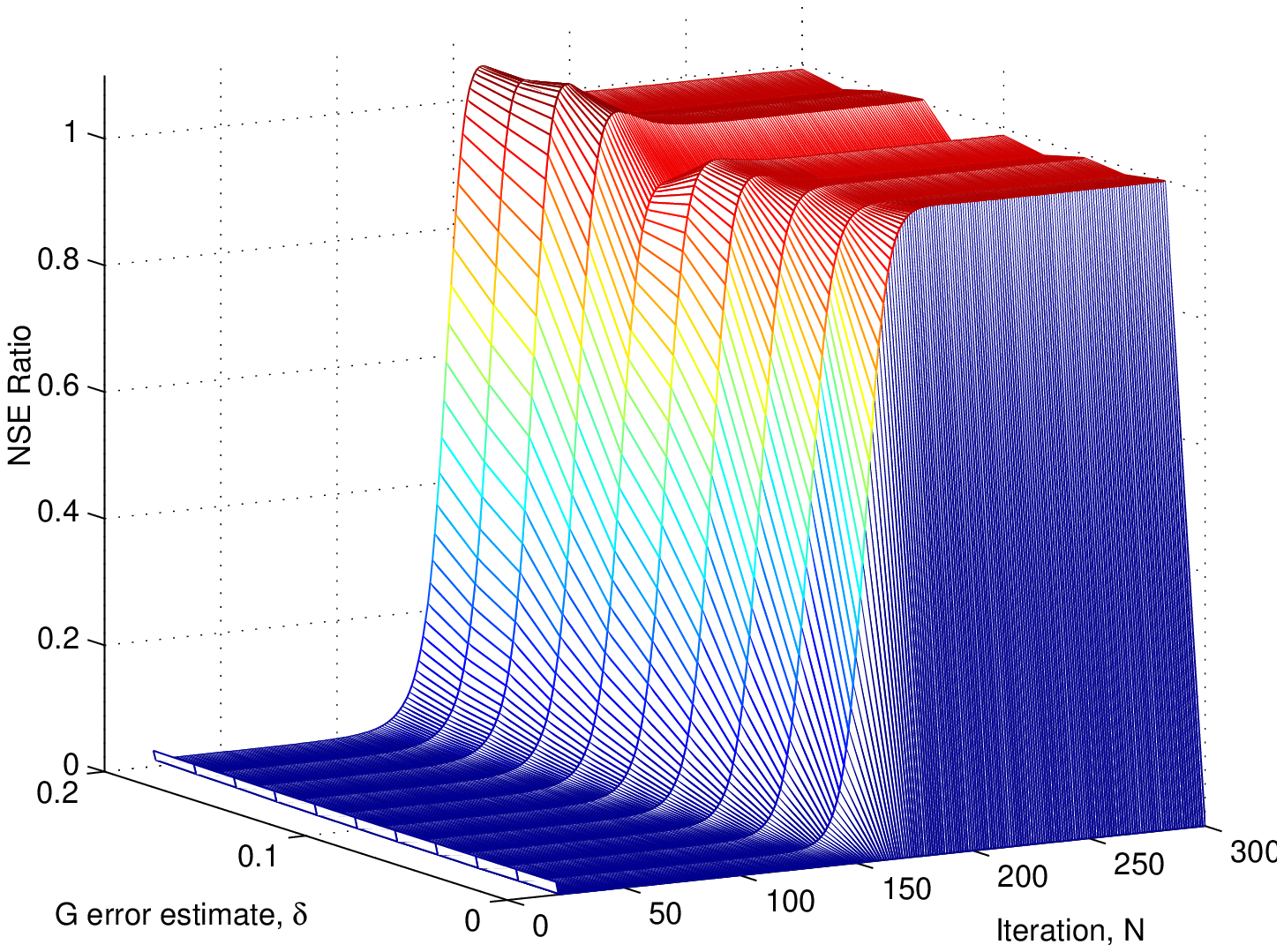}
\includegraphics[width=.47\textwidth]{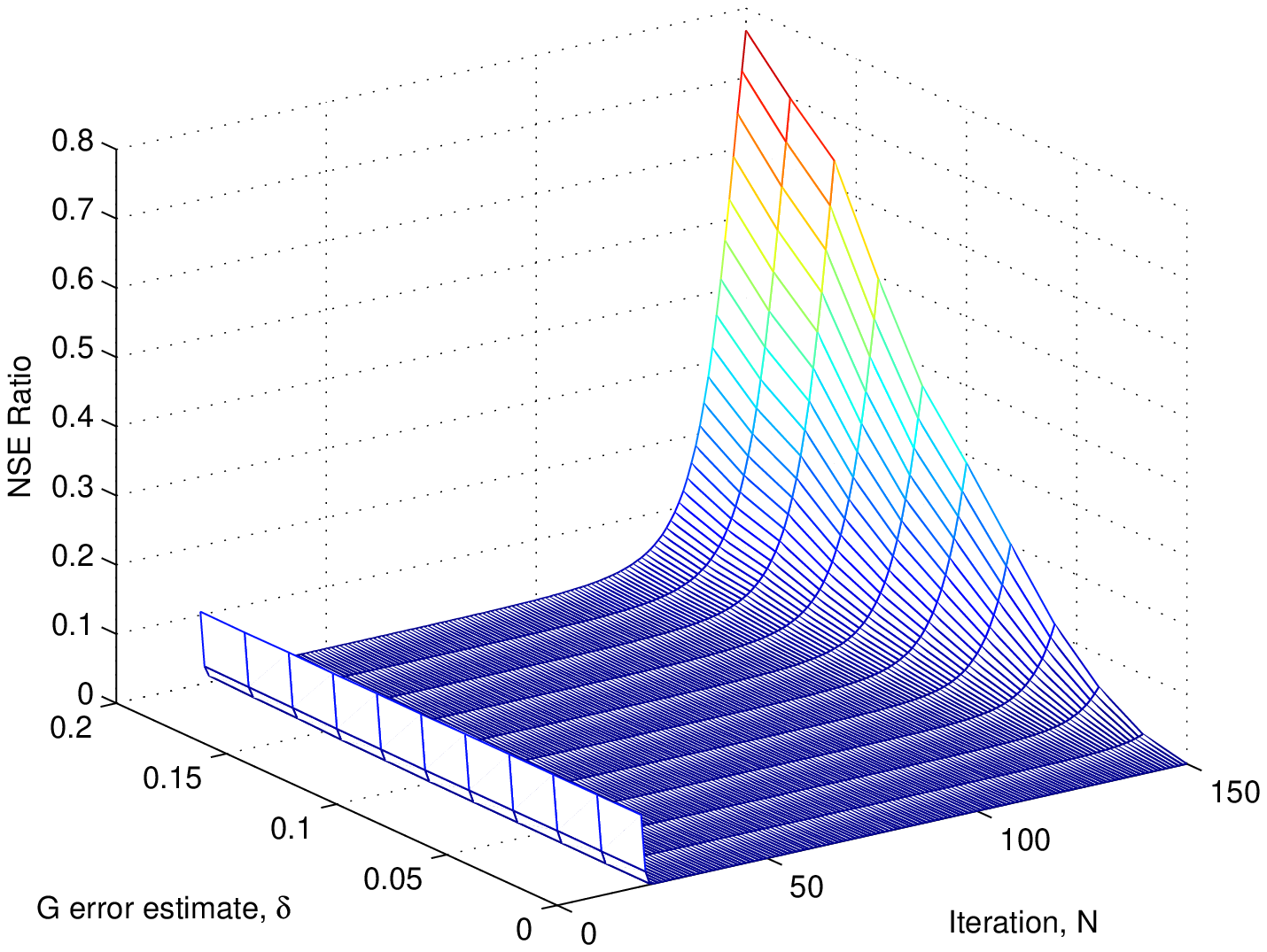}
\vspace{-4mm}
\caption{NSE Ratio considering fast ($\alpha=0.9$) and slow ($\alpha=0.1$) power convergence behavior of $K=7$ mt, and $\delta$ channel gain estimation error values. Bottom graph is a zoom in over $N\in[21;\,150]$ initial iterations.}
\label{fig:nser_K7_alpha0.1_0.9}
\end{figure}

Since we have some idea how fast the Verhulst algorithm reaches the equilibrium with different values of $\alpha$, it is important to determine solution quality in terms of convergence time. For the same system configuration of Fig. \ref{fig:converg_alpha0.1_0.9}, we have obtained in Fig. \ref{fig:nser_K7_alpha0.1_0.9} the associated NSE ratio, defined as:
\begin{equation}
    NSER =\frac{NSE (\alpha=0.9)}{NSE (\alpha=0.1)}=\frac{NSE ({\rm fast \,\, converg.})}{NSE ({\rm slow \,\, converg.})}
\end{equation}

One can see from Fig. \ref{fig:nser_K7_alpha0.1_0.9} that regardless of channel error estimates $\delta$, the quality of solution for both $\alpha = 0.9$ (fast) and $\alpha = 0.1$ (slow convergence) at the final section of iterations ($N>170$) shows high similarity ($NSER\approx 1$), but with a slight advantage in terms of convergence for $\alpha = 0.1$. In that region, with both convergence factors, the algorithm approaches to the optimal solution at same speed; as a consequence the $NSER \rightarrow 1$. Conversely, after a initial approaching convergence, i.e., after $23$ and until $\approx 120$ iterations, the Verhulst algorithm with $\alpha=0.9$ produces a much better solution, resulting in $NSE(\alpha=0.9) << NSE(\alpha=0.1)$. Due to the insufficient number of iterations, the algorithm is not able to achieve convergence for $\alpha=0.1$.

In conclusion, the best choice for $\alpha$ depends on the number of iterations constraint. If the number of iteration is a concern, the natural choice consists in to adjust the convergence factor as high as possible. Otherwise low values for $\alpha$ produce NSE slightly smaller.


\subsubsection{Solution Quality as a Function of System Loading}
\begin{figure}[!htbp]
\centering
\includegraphics[width=.45\textwidth]{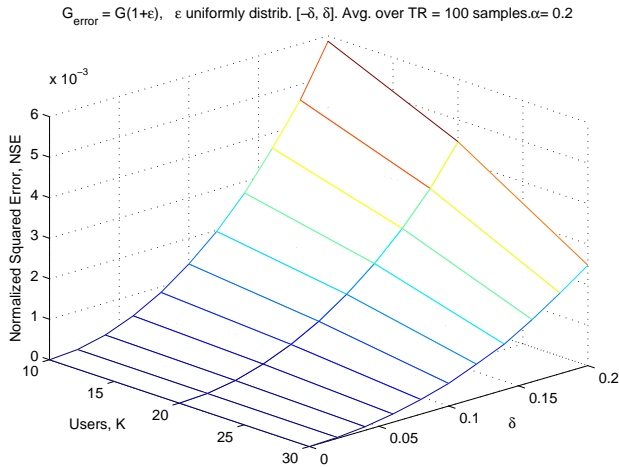}
\vspace{-4mm}
\caption{NSE degradation as a function of $K$ mobile terminals and $\delta$ channel gain estimation error, $\alpha=0.2$.}
\label{fig:nse_alpha0.2}
\end{figure}

Fig. \ref{fig:nse_alpha0.2} shows the average NSE behavior when the channel gain error $\delta$ increases for the $1000$th iteration and increasing system loading, $K = 10, 20$ and $30$ mobile terminals with different user class rates realizations (and uniformly distributed over the three user class rates). The convergence factor was assumed fixed $\alpha = 0.2$ and the algorithm convergence ran $100$ times to each combination of $K$, $\delta$ and user rates.

Note from Fig. \ref{fig:nse_alpha0.2} that the NSE values increase for low system loading (small $K$), showing an increasing degradation rate under specific system loading when the channel error estimates $\delta$ increase. We can explain this dependence by granularity effect, i.e., under high system loading, the average norm distances between the proposed algorithm solution and the analytical optimum solution results smaller due to large number of active users (high granularity), when compared to the low loading system cases (small $K$ and low granularity).  Anyway, in terms of NSE, the channel gain error estimates has a progressive effect over the solution quality.

\subsection{Adaptive Convergence Methods Performance}
In order to speed up the algorithm convergence, we have suggested in section \ref{sec:conv_speedup} two adaptive criteria based on SNIR's difference. There are two important performance aspects to be analyzed, considering adaptive methods against fixed $\alpha$ optimization methods: convergence time and solution quality. However, in order to privilege the quality solution analysis, in this subsection the lower convergence factor was adopted for the three methods; hence, in the next we evaluate the results just in terms of convergence time (number of iterations, $N$), considering fixed the convergence factor, $\alpha = 0.1$.

\begin{figure}[!htbp]
\centering
\includegraphics[width=.5\textwidth]{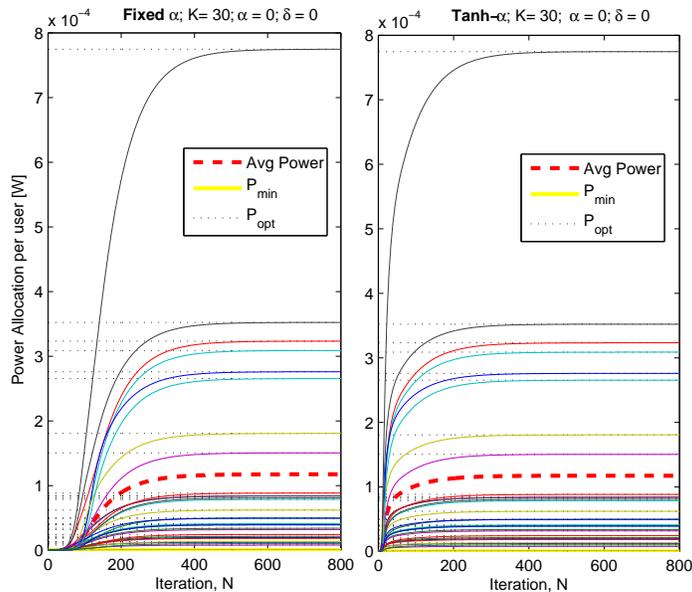}
\vspace{-7mm}
\caption{Convergence speed: adaptive $\alpha$ with tanh function (right) and the classical fixed $\alpha$ method (left).}
\label{fig:convergence_axf}
\end{figure}

A first approach to evaluate the reduction in the convergence time with the adoption of adaptive convergence factor is provided in Figure \ref{fig:convergence_axf}. Both plots were generated under the same channel conditions with $K = 30$ users and no channel gain error estimates. As expected, the convergence rate, mainly at the beginning (early iterations), is greatly speeded up.  One can see that roughly the adaptive tanh$-\alpha$ procedure allows the proposed algorithm to achieve total convergence $50\%$ early  regards to the fixed factor $\alpha = 0.1$.

\begin{figure}[htbp]
\centering
\includegraphics[width=.49\textwidth]{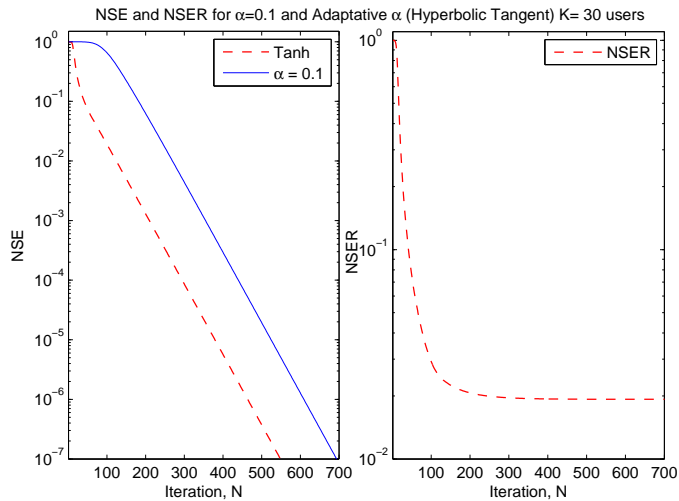}
\vspace{-4mm}
\caption{NSE and NSER to the adaptive method using tanh function against fixed $\alpha = 0.1$.}
\label{fig:nse_axf}
\end{figure}

In order to quantify the $\alpha$-adaptation effect over the normalized square error, Fig. \ref{fig:nse_axf} shows the NSE and NSER for each number of iterations in the range $[0; 700]$, and considering the same scenario of Fig. \ref{fig:convergence_axf}. Note that for any iteration after the initial iterations ($N>100$) the adaptive convergence factor provides at least one and half order better performance in terms of NSE. Specifically, for $N > 200$ results $NSE(\alpha_{\rm Adpt}) \approx2\cdot 10^{-2}NSE(\alpha = 0.1)$. In other words, the proposed tanh $\alpha$ adaptive method can achieve the same solution quality of $\alpha = 0.1$ using less iterations ($\approx 140$ less iterations when $N>100$).

\begin{figure}[htbp]
\centering
\includegraphics[width=.48\textwidth]{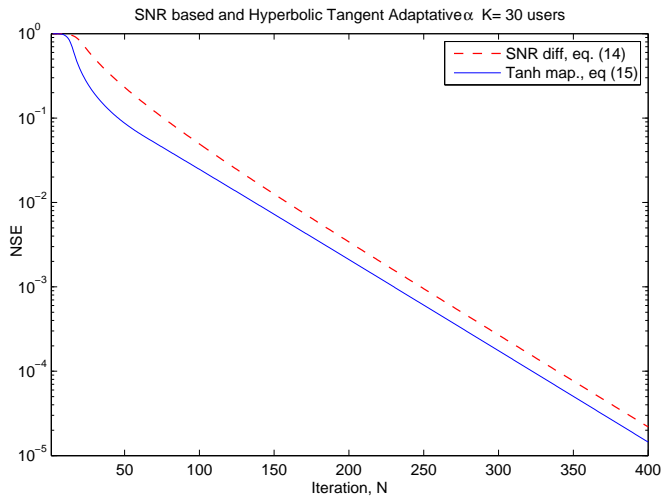}
\vspace{-4mm}
\caption{NSE for both $\alpha-$adaptive methods. $K=30$ users}
\label{fig:nse_axa}
\end{figure}

Finally, in order to determine the best adaptive convergence factor method suggested by (\ref{eq:adpt_method_a}) and (\ref{eq:adpt_method_b}) we can compare the NSE for both methods under the same channel and system conditions. Fig. \ref{fig:nse_axa} shows the simulation results considering $K = 30$ users and $\delta = 0$. Note that the hyperbolic tangent mapping always results a lower NSE, although this difference is marginal. Therefore, for any number of iterations the convergence solution provided by the $\tanh$ method is better than that provided by the SNR to SNR target difference mapping of (\ref{eq:adpt_method_a}),  at cost of a marginal increment in computational effort spent with tanh evaluation.

\subsection{Computational Complexity}
As in a distributed method, each link mobile terminal-BS performs separately their updating, i.e., as a whole the power control is performed by $K$ processors in parallel and each one performs only scalar operations. Hence, the analysis is reduced to the study of one iteration on each mobile terminal. So, comparing the proposed and the classical Foschini \cite{Foschini_93} algorithms, both result in same complexity.

On the other hand, in order to compare the computational complexity of the proposed algorithm with analytical matrix inversion approach, we have to quantify the number of additions and multiplications as a function of the number of interfering mobile terminals $(K-1)$. Equations (\ref{eq:Verhulst}), (\ref{newCIR}) and (\ref{eq:adpt_method_b}) are evaluated at each iteration in each terminal using the proposed algorithm with adaptive $\alpha$-tanh method. Table \ref{tab:complexity} shows the number of additions and multiplications operations executed per iteration. The $K$ tanh evaluations per iteration were admitted as a look-up table operations.
\begin{table}[!h]
\centering
\caption{Operations per iteration for the proposed algorithm executed on each mobile terminal.} \label{tab:complexity}
\vspace{-.3cm}
\small
\begin{tabular}{llc}
\hline
Equation & Operation & Number of Operations \\
\hline\hline
(\ref{eq:Verhulst}) &Additions & $2$ \\
	                &Multiplications & $3$ \\
\hline
(\ref{newCIR}) & Additions & $K$ \\
	           & Multiplications & $K+3$ \\
\hline
(\ref{eq:adpt_method_b}) & Additions & $1$ \\
                         & look-up table &$1$\\
\hline
\end{tabular}
\end{table}

Hence, computational complexity of the proposed algorithm is $\mathcal{N}(K+10)$, where $\mathcal{N}$ is the number of iterations necessary for convergence. Comparing with the best case complexity of the matrix inversion operation, which is given by $\mathcal{O}[K^2 \, \cdot \log(K)]$ \cite{Golub96, Amund03}, the proposed optimization methodology achieves a considerable complexity reduction when the number of mobile terminals is large and the NSE requirement is not excessively tight. Besides, in the proposed method, the complexity could be controlled simply specifying the maximal admissible NSE.

\section{Conclusions} \label{sec:concl}
An extension on discrete Verhulst power equilibrium approach, previously suggested in literature was proposed, taking into account the jointly power-rate optimal allocation problem. For this purpose, multirate users associated to voice, data and video types of traffic were aggregated as distinct user' classes, with the assurance of QoS and minimum rate allocation per user. Furthermore, two criteria for convergence speed up were suggested and compared with the fixed convergence factor case.

Numerical results for convergence time (number of iterations), quality of solution (NSE) and number of basic operations (multiplications and sums) point out advantages of the Verhulst power-rate $\alpha-$adaptive algorithm  when compared to analytical solution based on the interference matrix inverse.

Finally, the logistic map approach applied to resource allocation problem in DS/CDMA systems suggested here demonstrates tremendous potential of applicability. Future directions include a) power-rate allocation for multiple-input-multiple-output (MIMO) CDMA systems, and b) discrete Verhulst equilibrium adaptation to jointly minimize power consumption and maximize the throughput by changing the last constraint in (\ref{eq:Equilibrium}) to $R^\ell \geq R_{\min}^\ell$.


\end{document}